\documentclass[sn-mathphys,Numbered]{sn-jnl}
\usepackage{graphicx}%
\usepackage{multirow}%
\usepackage{amsmath,amssymb,amsfonts}%
\usepackage{amsthm}%
\usepackage{mathrsfs}%
\usepackage[title]{appendix}%
\usepackage{xcolor}%
\usepackage{textcomp}%
\usepackage{manyfoot}%
\usepackage{booktabs}%
\usepackage{algorithm}%
\usepackage{algorithmicx}%
\usepackage{algpseudocode}%
\usepackage{listings}%
\begin{document}
\title[Article Title]{Assessing Lepton Flavor Universality Violations in Semileptonic Decays}
\author[1]{\fnm{Sonali} \sur{Patnaik}}\email{sonali\_patnaik@niser.ac.in}
\author[1]{\fnm{Lopamudra} \sur{Nayak}}\email{lopalmnayak@niser.ac.in}
\author*[1,2]{\fnm{Rajeev} \sur{Singh}}\email{rajeevofficial24@gmail.com}
\affil[1]{\orgdiv{School of Physical Sciences}, \orgname{National Institute of Science Education and Research, An OCC of Homi Bhabha National Institute}, \orgaddress{\city{Jatni}, \postcode{752050}, \state{Odisha}, \country{India}}}
\affil[2]{\orgdiv{Center for Nuclear Theory, Department of Physics and Astronomy}, \orgname{Stony Brook University}, \orgaddress{\city{Stony Brook}, \postcode{11794-3800}, \state{New York}, \country{USA}}}
\abstract{In light of recent measurements suggesting potential lepton flavor universality violations in semileptonic decays at collider experiments, this article provides a concise study of tree- and loop-level $B$-hadron semileptonic decays, $b \to c l \nu_l$ and $b \to s l^+ l^-$. We provide predictions for lepton flavor violating observables, $\mathcal{R}_{J/\psi}$ and $\mathcal{R}_{\eta_c}$, across the entire $q^2$ range. Our study employs the Relativistic Independent Quark Model (RIQM), highlighting a model-dependent approach to these observables. We compare our model's predictions with existing lattice predictions, demonstrating the strong applicability of the RIQM framework in describing $B_c$ decays. Additionally, we reassess global averages for $\mathcal{R}_{D(D^*)}$ and $\mathcal{R}_{K(K^*)}$ in semileptonic transitions. With the upcoming experimental upgrades and the anticipated Run 3 data on $B_c$ meson decays, rapid confirmation of these quantities could indicate significant evidence of physics beyond the Standard Model, thereby opening new pathways for understanding the complex flavor dynamics in $B$ meson decays.}



\maketitle
\section{Introduction}
\label{sec:intro}
The Standard Model (SM) of particle physics identifies the basic elements as fermions, including quarks and leptons, as detailed in Ref.~\cite{ParticleDataGroup:2022pth}. Leptons, considered as point-like particles without substructure, partake in weak and electromagnetic interactions, while neutrinos are involved solely in weak interactions. The SM recognizes these charged leptons ($e^-$, $\mu^-$, $\tau^-$) as universal across generations, except for variations due to their differing masses, a concept validated in multiple decay experiments~\cite{BESIII:2013csc,NA62:2012lny,PiENu:2015seu,ParticleDataGroup:2022pth,ALEPH:2005ab}. 
The Lepton Flavor Universality (LFU), an accidental symmetry in the SM, results from varying Yukawa interactions between lepton-Higgs leading to different lepton masses ($m_{\tau}>m_{\mu}>m_e$)~\cite{HFLAV:2016hnz,Bifani:2018zmi,Gambino:2020jvv,Bernlochner:2021vlv}. However, recent studies suggest potential LFU breaking in $b \to s l^+ l^-$ $(l = e, \mu)$, challenging our fundamental understanding of physics and implying the possible existence of new particles that violate this symmetry~\cite{LHCb:2021trn}. 
These could alter the rates of quark-lepton transitions in the $B$-hadron decays of the SM~\cite{Bonilla:2022qgm,Gedeonova:2022iac,Castro:2022qkg,Descotes-Genon:2022qce,Buras:2022qip,Datta:2022zng,Barbosa:2022mmw,Dutta:2017xmj,Dutta:2018vgu,Dutta:2019wxo,Mohapatra:2021ynn,Dutta:2013qaa,Rajeev:2020aut,Rajeev:2021ntt,Das:2021lws,Das:2023gfz,Gao:2021sav,Cui:2023bzr,Huang:2018nnq}. Anomalies hinting at LFU violation have also been observed in the first-order decays of beauty ($B$) mesons to final $\tau$ lepton states~\cite{BaBar:2012obs,BaBar:2013mob,Belle:2009zue, Belle:2015qfa, Belle:2016dyj,BELLE:2019xld,Belle:2019oag,LHCb:2015gmp,LHCb:2016ykl,LHCb:2017avl,LHCb:2019efc,LHCb:2021lvy,LHCb:2021trn,LHCb:2021xxq, LHCb:2021zwz}. 

Additionally, leptonic and semileptonic decays of $B$ mesons appear to contest lepton universality~\cite{Belle:2009zue,BaBar:2012mrf,LHCb:2014cxe,LHCb:2014vgu,LHCb:2015tgy,LHCb:2015wdu,LHCb:2015svh,Belle:2016fev,LHCb:2016ykl,CMS:2017rzx,LHCb:2017avl,ATLAS:2018gqc}. Detecting such violations and elucidating the interactions of new particles could pave the way for investigating physics beyond the SM (BSM), potentially unveiling insights into dark matter, matter-antimatter asymmetry, and the dynamics of the electroweak scale.

This paper reviews anomalies in semileptonic transitions, focusing on differences in $B$ decay rates to $\tau$ and $\mu$ leptons, stemming from the notable $\mu$-$\tau$ mass disparity. The first charge-current anomalies were observed in $B \to D, D^* l {\bar\nu_l}$ decays by the BaBar Collaboration~\cite{BaBar:2012obs}, defining the LFU ratio as
\begin{equation}
\mathcal{R}_{D^{(*)}}=\frac{Br(B \to D^{(*)} \tau {\bar\nu_{\tau}})}{Br(B \to D^{(*)} l {\bar\nu_l})}\,, \quad \text{with}\quad \ell= \mu,e\,,
\end{equation}
where $D^{(*)}$ denotes $D(D^*)$ meson, and $Br$ represents the branching ratio. The results deviated from the SM by 3.4\,$\sigma$~\cite{BaBar:2012obs} and were later affirmed by measurements at BaBar~\cite{BaBar:2013mob}, Belle~\cite{Belle:2015qfa}, and LHCb~\cite{LHCb:2015gmp}.
Recently, LHCb provided an updated measurement of $\mathcal{R}_D$ and $\mathcal{R}_{D^*}$, based on LHC Run 1 data, aligning with previous findings~\cite{HFLAV:2019otj,HFLAV:2022pwe,averages:2022spring,LHCb:2023cjr}. This $\tau$ reconstruction in $\tau \to \mu \nu {\bar \nu}$ supersedes the 2015 result~\cite{LHCb:2015gmp}, with measured observables exceeding SM predictions, implying Lepton Flavor Universality Violation (LFUV). While recent averages edge closer to SM predictions~\cite{Iguro:2020cpg, Bordone:2019vic}, the deviation significance remains above 3\,$\sigma$ due to reduced uncertainties. These theoretical and experimental measurements promote further exploration of form factors in semileptonic transitions.
The study of $\mathcal{R}$ in various  semileptonic decay modes prompted LHCb's first results for $\mathcal{R}(J/\psi)$ from $B_c \to  J/\psi$ decay, defined as
\begin{equation}
\mathcal{R}_{J/\psi}=\frac{Br(B_c{^+} \to J/\psi \tau^+ \nu_{\tau})}{Br(B_c{^+} \to J/\psi \mu^+ \nu_{\mu})}\,,
\end{equation}
with $\tau^+$ decaying leptonically to $\mu^+$ $\nu_{\mu}$ ${\bar \nu_{\tau}}$. This analysis utilized a 3fb$^{-1}$ pp collision data sample at $\sqrt{s}=7$ and $8$ TeV~\cite{LHCb:2017vlu}. The $\tau$ lepton was reconstructed, and a global fit performed, revealing consistent deviation from SM predictions, currently between 0.25-0.28~\cite{Ivanov:2006ib,Hernandez:2006gt,Watanabe:2017mip}, roughly 2\,$\sigma$ lower. These deviations hint at LFU violation.

Further LFU tests are underway at LHCb for other $B_q$ species, including $\mathcal{R}(D^+)$ and $\mathcal{R}_{{\Lambda_c}^*}$~\cite{Hernandez:2006gt,Ivanov:2006ni}. LFU measurements in $b \to s l^+ l^-$ transitions present anomalies~\cite{LHCb:2014vgu,LHCb:2017avl, LHCb:2019hip, LHCb:2019efc, BELLE:2019xld, LHCb:2021trn, LHCb:2021lvy}, and LFUV observables
\begin{equation}
{\mathcal {R}}_{K^{(*)}}=\frac{Br(B \to K^{(*)} \mu^+ \mu^-)}{Br(B \to K^{(*)} e^+ e^-)}\,, \quad {\cal R}_{\phi} = \frac{Br(B_s \to \phi \mu^+ \mu^-)}{Br(B_s \to \phi  e^+ e^-)}\,,
\end{equation}
display deviations from SM predictions, significant to 3.1$\sigma$, with a combined uncertainty of 5\%~\cite{Bordone:2016gaq,Isidori:2020acz}. LHCb's recent tests on muon-electron universality in $B^+ \to K l^+ l^-$ and $B^0 \to K^0 l^+ l^-$ decays, using $B$ meson data from Run 1 and 2 (total luminosity: 9 fb$^{-1}$), concur with SM predictions~\cite{LHCb:2022qnv}. Updated measurements for $\mathcal{R}_K$ and $\mathcal{R}_{K^*}$ supersede previous ones~\cite{LHCb:2021trn} and align with SM values. We reassess these transitions while constructing observables and analyzing their SM tension. Table~\ref{table:3p} contains the summary of experimental world averages versus SM predictions.
\begin{table*}[!hbt]
\centering
\setlength\tabcolsep{10pt}
\renewcommand{\arraystretch}{1.5}
\caption{Summary of experimental measurements versus SM predictions~\cite{Patnaik:2022moy}}
\resizebox{\textwidth}{!}{
\begin{tabular}{|c|c|c|c|}
\hline LFU Parameters&LHCb measurements&SM prediction&Deviation.\\
\hline${\mathcal {R}}_D^{LHCb2022}$&0.441 $\pm$ 0.060 $\pm$ 0.066~\cite{Iguro:2022yzr}&0.298 $\pm$ 0.004~\cite{HFLAV:2022pwe}& 2.16\,$\sigma$\\
\hline${\mathcal {R}}_{D^*}^{LHCb2023}$&0.257 $\pm$ 0.012 $\pm$ 0.014~\cite{LHCb:2023cjr} &0.254 $\pm$ 0.005~\cite{HFLAV:2022pwe}&1\,$\sigma$ \\
\hline${\mathcal {R}}_{J/\psi}$&0.71 $\pm$ 0.17 $\pm$ 0.181~\cite{LHCb:2017vlu}&0.283 $\pm$ 0.048~\cite{Watanabe:2017mip}&2\,$\sigma$\\
\hline${\mathcal {R}}_K^*$&1.027 $\pm$ 0.072 $\pm$ 0.027~\cite{LHCb:2017vlu}&1.00 $\pm$ 0.0.01~\cite{Bordone:2016gaq}&0.2\,$\sigma$\\
\hline${\mathcal {R}}_K$&0.949$\pm$ 0.042$\pm$ 0.022~\cite{LHCb:2017vlu}&1.00 $\pm$ 0.01~\cite{Bordone:2016gaq}&0.2\,$\sigma$\\
\hline
\end{tabular}}
\label{table:3p}
\end{table*}
Quantum Chromodynamics (QCD) addresses quark confinement inside hadrons and characterizes decay transition rates through Lorentz invariant form factors. These factors represent information about hadrons as bound systems. Yet, QCD complexities from non-abelian and non-perturbative properties hinder the extraction of these factors. Instead, phenomenological models are employed to describe the bound state nature of hadrons and decay properties. Various theoretical methods, such as heavy quark effective theory (HQET), quark models, and Lattice QCD (LQCD), aim to explain $B$-decay anomalies~\cite{Crivellin:2017zlb, Crivellin:2018yvo, Crivellin:2019dwb, Blanke:2018sro,Calibbi:2017qbu,Calibbi:2015kma}.
These methods have strengths and weaknesses. Regardless of their Lorentz interaction potential structure, a quark potential model's success is determined by its ability to reproduce observed data across various hadron sectors. The model's reliability is based on its accurate depiction of constituent-level dynamics inside the hadron core and predictions of various hadronic properties, including decays. The downside is that potential models are not unique and may only reproduce experimental data within a limited range. Hence, expanding a quark model's applicability to a broader range of observed data is crucial.

In this work, we present our findings on $B_c$-decay anomalies within the Relativistic Independent Quark Model (RIQM) framework, aiming to test the model's ability to explain LFU ratios. Given the availability of SM LQCD vector and axial form factors for the $B_c \to J/\psi$ decay, which have motivated us to further undertake this study and therefore we present a systematic comparison of our results with LQCD inputs.
This paper is organised as follows: Section~\ref{sec:exp_outlook} presents an experimental outlook on LFU. Sections~\ref{sec:testsofLFU} and \ref{sec:tests} discuss model-dependent studies of $b\to c$ and a brief review on $b\to s$ decays, respectively. The final section~\ref{sec:conclusions}, offers our conclusions and future directions.
\section{Experimental outlook on Lepton Flavor Universality}
\label{sec:exp_outlook}
The discovery of the $b$ quark~\cite{E288:1977xhf} paved the way for the production of $B$-hadrons at colliders such as CESR, LEP, and Tevatron. Comprehensive studies on third-generation LFUV in $B$ mesons only became viable with the advent of the $B$-factories and the LHC~\cite{LHCb:2008vvz}. 
The examination of $B$ meson decay is split into two: decays that involve FCNC (flavor-changing neutral current) ($b$-quark to $s$-quark transition with lepton pair emission) and FCCC (flavor-changing charge current) decays (transition from a $b$-quark to a $c$-quark, with leptons and neutrinos emission). New physics (NP) mediators like leptoquarks~\cite{Hiller:2014yaa,Gripaios:2014tna,deMedeirosVarzielas:2015yxm,Barbieri:2016las} and $Z^\prime$~\cite{Altmannshofer:2014cfa,Crivellin:2015mga,Celis:2015ara,Falkowski:2015zwa} significantly modify their amplitudes, facilitating their study.
Since 2010, LHCb has collected a large data set of $b \bar b$ pairs, compensating for the challenging pp collision environment \cite{LHCb:2008vvz}. The detector design optimizes heavy meson decay studies, with excellent identification and momentum resolution capabilities \cite{LHCb:2014set}. The LHCb platform also allows for LFU tests on all $B$-hadron species produced at the LHC, including those with low production rates like $B_c$ mesons \cite{LHCb:2014mvo}. 
The Belle detector is also instrumental in heavy meson decay studies and is equipped with various high-precision components. Alongside the upgraded Belle II \cite{Belle:2009zue,Bernlochner:2021vlv}, they have already made significant strides in the examination of $B \to D(D^*) l \nu_l$ decays \cite{BaBar:2012obs,BaBar:2013mob,Belle:2015qfa,Belle:2016kgw,Belle:2016dyj,LHCb:2015gmp,LHCb:2017smo}. 
Belle II and the upgraded LHCb detector are set to continue data collection for the next 15 years, anticipated to multiply current data samples. Measurements from Belle II and LHCb are expected to illuminate current flavor anomalies and align with SM predictions \cite{Albrecht:2017odf,Belle-II:2020sdf, Belle-II:2022cgf}.
Experiments at ATLAS \cite{ATLAS:2008xda} and CMS \cite{CMS:2008xjf} also contribute significantly to our understanding of physics beyond the SM. Both ATLAS and CMS, with the CMS having recently made a significant detection of an excited $B_c$ state \cite{CMS:2019uhm}, are limited compared to LHCb in flavor physics capabilities due to certain requirements and limitations but provide complementary contributions to muon decay processes.

The rare decay of $B_s \to \mu^+ \mu^-$ has been studied extensively by CMS using data from LHC Run 2. This study revealed a clear signal of $B_s$ meson decaying to a muon-antimuon pair, the decay rate of which closely matches theoretical predictions. This result, alongside future studies of extremely rare transitions like $B^0 \to \mu^- \mu^+$ decay expected to be facilitated by data from LHC Run 3, could prove instrumental in understanding puzzling anomalies and the very physics beyond the standard model.
\section{Tests of LFU ratios in \texorpdfstring{$B_c\to X l \nu_l$ where $X$ is $\eta_c$, $J/\psi$, $D$ and $D^*$ and $l = e, \mu, \tau$}{}}
\label{sec:testsofLFU}
\begin{figure}[h!]
\begin{center}
\includegraphics[width=100mm]{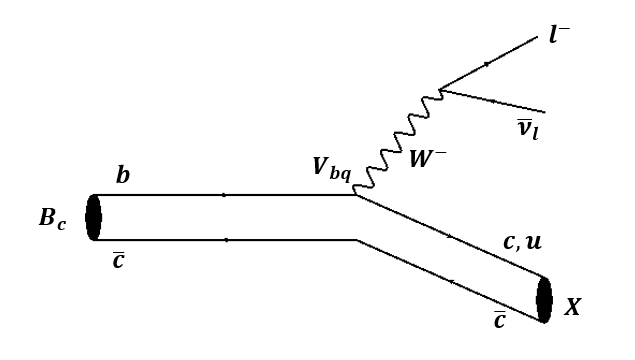}
\end{center}
\caption{SM contribution for $B_c \to X l \nu_l$ where $X = \eta_c , J/\psi$ , $D(D^*)$~\cite{Patnaik:2022moy}.}	
\label{fig:3}
 \end{figure}

In this section, we explore the ratios of $b \to \tau, \mu$ leptons due to their mass difference influencing decay amplitudes. Our focus is on the SM tree-level depiction of $B_c \to X l \nu_l$, where $X = \eta_c , J/\psi$ , $D(D^*)$, as shown in Fig.~\ref{fig:3}. We examine the $B_c$ meson for its unique traits as the only bound state of two heavy open flavored quarks (charm and bottom). Its position between charmonium $(c \bar c)$ and bottomonium $(b \bar b)$ makes it intriguing but not well-studied due to limited data. It decays exclusively via weak interaction as it resides below the $B\bar D$ threshold, yielding a long life span and multiple weak decay channels with notable branching ratios.

Recent evidence of an excited $B_c$ state was found by the CMS Collaboration~\cite{CMS:2019uhm} through the study of  $B_c^+\to\pi^+\pi^-$ with a total integrated luminosity of 143 fb$^{-1}$ at $\sqrt{s} =$ 13 TeV, determining a $B_c$(2S) meson mass of $6871\pm 1.2\pm 0.8$ MeV. Detection of ground and excited states of $B_c^*$ remains elusive. However, the energy and higher luminosity at LHC and $Z_0$ factory should bolster the accumulation rate of these undetected states, promoting a deeper understanding of $B_c$ and $B_c^*$ counterparts.
This emerging data and the anticipated surge of $B_c$ events in ongoing and forthcoming experiments underpin the importance of investigating decay properties in this area. Thus, the $B_c$-meson offers a unique view of heavy quark dynamics and a distinct test for QCD.
In light of these points, we offer a model-dependent discussion to reconcile the disparities between SM and BSM physics.
\subsection{RIQM framework}
Studying exclusive semileptonic decays involving non-perturbative hadronic matrix elements poses challenges due to current limitations in rigorously applying QCD principles for reliable invariant transition amplitude measurements. As a result, various theoretical approaches utilize phenomenological models to investigate non-perturbative QCD dynamics~\cite{AbdEl-Hady:1999jux,Hernandez:2006gt,Ivanov:2006ib,Colquhoun:2015oha}. HQET's usefulness lies in relating form factors via $1/m_Q$ and $\alpha_s$, and several quark models like QCD Sum rule, light cone sum rule, and lattice QCD are employed to approximate these form factors. Form factor parametrization insights shape differential decay amplitudes and provide measurements on NP scales.
We thus present an overview of global averages in a model-dependent approach. We used the RIQM model, based on confining harmonic potential in an equally mixed scalar-vector form~\cite{Patnaik:2019jho,Patnaik:2023efe}
\begin{equation}
U(r)=\frac{1}{2}\left(1+\gamma^0\right)\,V(r)\,.
\label{eq:harmonicpotential}
\end{equation}
In the potential $V(r) = (ar^2 + V_0)$, $r$ is the quark-antiquark relative distance, $\gamma^0$ is the time-like Hermitian matrix, and $a$ and $V_0$ are potential parameters. The values of the potential parameters is listed in \ref{subsec:Num} which are set from prior model application. The internal dynamics of quarks are represented by a quark Lagrangian density of suitable Lorentz structure
\begin{equation}
{\cal L}^{0}_{q}(x)={\bar \psi}_{q}(x)\;\left[\frac{i}{2}\,\gamma^{\mu}
\partial _\mu - m_q-U(r)\;\right]\;\psi _{q}(x)\,.
\end{equation}
This leads to the individual quark's Dirac equation
\begin{equation}
\left[\gamma^0E_q-{\vec \gamma}.{\vec p}-m_q-U(r)\right]\psi_q(\vec r) = 0\,,
\end{equation}
where $\psi_q(\vec r)$ is the Dirac normalized wave function, expressible in a two-component form
\begin{eqnarray}
\psi_q(\vec r)=\left(
\begin{array}{c}
\psi_A(\vec r)\\
\psi_B(\vec r)
\end{array}\;\right)=\left(
\begin{array}{c}
\psi_A(\vec r)\\
\left (\frac{{\vec \sigma}\cdot{\vec p}}{E_q+m_q}\right)\;\psi_A(\vec r)
\end{array}\right)\,.
\end{eqnarray}
In this form, $\psi_A(\vec r)$ and $\psi_B(\vec r)$ represent the quark wave function's upper and lower components of opposite parity, respectively.

The RIQM, like other potential models, is a QCD-inspired model that determines the confinement of quarks in a hadron through an interaction potential with appropriate Lorentz structure. Theoretical derivations of observable properties of composite hadrons from their constituent dynamics should be feasible but have proven challenging due to inherent QCD complexities. As such, numerous effective quark potential models have been used as phenomenological means to describe these dynamics. 
The model assumes a non-perturbative multi-gluon interaction represented by the potential form, with residual interactions arising from quark-pion coupling and short-distance one-gluon exchange treated perturbatively. The confining potential $U(r)$, as outlined in Eq.~\eqref{eq:harmonicpotential}, provides a simple yet effective framework for analyzing various hadronic properties, including FCCC-involved decays.
We aim to broaden the applicability of the RIQM, showcasing its suitability as an alternative phenomenological scheme for examining various hadronic phenomena, and comparing results with other theories and available experimental data. The model's efficacy has been verified across a range of hadronic phenomena, including various types of hadron decays~\cite{Barik:1992pq,Barik:1994vd,Priyadarsini:2016tiu,Barik:1995sq,Barik:1996kn,Barik:2001gr,Barik:1993yj,Barik:1993aw,Barik:1996xf,Barik:1997qq,Barik:2009zz,Barik:2008zza,Barik:2008zz,Barik:2009zza,Barik:2001vp,Barik:2009zzb,Naimuddin:2012dy,Kar:2013fna,Patnaik:2017cbl,Patnaik:2018sym,Patnaik:2019jho}. The invariant matrix element for $B_c\to \eta_c(J/\psi)l^-\bar{\nu}_l$ and $B_c\to D(D^*)l^-\bar{\nu}_l$ is expressed in general form as~\cite{Nayak:2021djn}
\begin{equation}
{\cal M}(p,k,k_l,k_\nu)={\frac{\cal G_F}{\sqrt{2}}}V_{bq^{'}}\,{\cal H}_\mu(p,k) \,{\cal L}^\mu(k_l,k_\nu)\,.
\end{equation}
The effective Fermi coupling constant is represented by ${\cal G}_F$, with $V_{bq^{'}}$ as the corresponding CKM parameter. ${\cal L}^\mu$ and ${\cal H}_\mu$ denote the leptonic and hadronic currents, respectively. The four-momentum of parent ($B_c$) and daughter (X) mesons, lepton, and neutrino are symbolized by $p,k,k_l,k_\nu$ respectively. It's important to note that decay processes occur when the involved mesons are in their momentum eigenstates. Thus, any field-theoretic decay process representation necessitates the depiction of meson bound states via suitable momentum wave packets, which illustrate the momentum and spin distribution between the constituent quark and antiquark within the meson core.

In the RIQM method, a wave packet representing a meson bound state, such as $\vert B_c(\vec{p}, S_{B_c})\rangle$, possessing a definite momentum $\vec{p}$ and spin $S_{B_c}$, is formulated as follows~\cite{Patnaik:2017cbl,Patnaik:2018sym, Patnaik:2019jho, Priyadarsini:2016tiu, Barik:1992pq, Barik:1993aw, Barik:1993yj, Barik:1994vd, Barik:1995sq, Barik:1996kn, Barik:1996xf, Barik:1997qq, Barik:2001gr, Barik:2001vp, Barik:2008zz, Barik:2008zza, Barik:2009zz}
\begin{equation}
\big\vert B_c \left(\vec{p},S_{B_c}\right)\big\rangle = \hat{\Lambda}\left(\vec{p},S_{B_c}\right)\big\vert \left(\vec{p_b},\lambda_b\right);\left(\vec{p_c},\lambda_c\right)\big\rangle \,.
\end{equation}
Here, $\vert (\vec{p_b},\lambda_b);(\vec{p_c},\lambda_c)\rangle $ denotes the Fock space representation of unbound quark and antiquark in a color-singlet setup, each with its momentum and spin. Meanwhile, ${\hat \Lambda}(\vec {p},S_{B_c})$ signifies an integral operator reflecting the meson's bound state properties.
\subsection{Helicity amplitudes and observables}
Using the model's dynamics, the weak form factors are integrated and the angular decay distribution in $q^2$ (where $q = p-k = k_l + k_\nu$) is expressed as~\cite{Nayak:2021djn}
\begin{equation}
\frac{d\Gamma}{dq^2 d\cos\theta} = {\frac{{\cal G}_F}{(2\pi)^3}}|V_{bq}|^2\frac{(q^2-m_l^2)^2}{8M^2 q^2}|\vec{k}|{\cal L}^{\mu \sigma}{\cal H}_{\mu \sigma}\,.
\label{eq:10}
\end{equation}
Here, ${\cal L}^{\mu \sigma}$ and ${\cal H}_{\mu \sigma}$ represent lepton and hadron correlation functions, respectively. The mass of the charged lepton is denoted as $m_l$, and $M$ is the mass of the parent ($B_c$) meson. 
The completeness property allows rewriting the lepton and hadron tensors in Eq.~\eqref{eq:10} as
\begin{eqnarray}
{\cal L}^{\mu \sigma}{\cal H}_{\mu\sigma} &=&{\cal L}_{\mu'\sigma'}g^{\mu'\mu}g^{\sigma'\sigma}{\cal H}_{\mu\sigma}\,,\nonumber\\
&=&{{\cal L}_{\mu'\sigma'}}{\epsilon^{\mu^{'}}}(m){\epsilon^{\mu^{\dagger}}}(m^{'}){g_{mm^{'}}}{\epsilon^{\sigma^{\dagger}}}(n){\epsilon^{\sigma^{'}}}(n^{'})g_{nn^{'}}{\cal H}_{\mu\sigma}\,,\nonumber\nonumber\\
&=&{L(m,n)}{g_{mm'}}{g_{nn'}}H({m'}{n'})\,.
\end{eqnarray}
The lepton and hadron tensors, defined in terms of helicity components, are as follows
\begin{eqnarray}
L(m,n)={\epsilon^{\mu}}(m){\epsilon^{\sigma ^\dagger}}(n){\cal L}_{\mu \nu}\,,\qquad
H(m,n)={\epsilon^{\mu^\dagger}}(m){\epsilon^{\sigma}(n)}{\cal H}_{\mu \nu}\,.
\end{eqnarray}
Expressing physical observables on a helicity basis simplifies the calculation. Hence, the helicity form factors are written as Lorentz invariant form factors that depict decay amplitudes. The Lorentz contraction in Eq.~\eqref{eq:10} can thus be performed with the helicity amplitudes, as illustrated in~\cite{Nayak:2021djn}.
In this study, we don't account for the azimuthal $\chi$ distribution of the lepton pair. Therefore, by integrating over the lepton tensor's azimuthal angle dependence, we obtain the differential partial helicity rates $(d\Gamma_i/dq^2)$
\begin{equation}
\frac{d\Gamma_i}{dq^2}=\frac{{\cal G}_f^2}{(2\pi)^3}{\vert V_{bq^{'}}\vert}^2 \frac{({q^2}-{m_l^2})^2}{12{M^2}q^2}|\vec{k}|H_i\,.
\end{equation}
Here, $H_i(i=U,L,P,S,SL)$ represents a standard set of helicity structure function given by linear combinations of helicity components of hadron tensor $H(m,n)=H_m H_n^\dagger$
\begin{eqnarray}
{H_U}&=&Re({H_+}{H_+^\dagger})+Re({H_-}{H_-^\dagger})\quad   :Unpolarized-transversed\,,\nonumber\\
{H_L}&=& Re({H_0}{H_0^\dagger})\quad \quad \quad \quad \quad \quad \quad \,\,\,\, :Longitudinal\,,\nonumber\\
{H_P}&=&Re({H_+}{H_+^\dagger})-Re({H_-}{H_-^\dagger})\quad :Parity-odd\,,\nonumber\\
{H_S}&=&3Re({H_t}{H_t^\dagger})\quad \quad \quad \quad \quad \quad \quad \,\,:Scalar\,,\nonumber\\
{H_{SL}}&=&Re({H_t}{H_0^\dagger})\quad \quad \quad \quad \quad \quad \quad \,\,\,\,\, :Scalar-Longitudinal\ Interference\,.\nonumber
\end{eqnarray}
 \begin{table}[!hbt]
 \renewcommand{\arraystretch}{1}
 \centering
 \setlength\tabcolsep{0.5pt}
 \caption{Helicity rates (in $10^{-15}$ GeV) of semileptonic $B_c$-meson decays into charmonium and charm-meson state~\cite{Patnaik:2022moy}}
 \label{tab1}
 \begin{tabular}{|c|c|c|c|c|c|c|c|c|c|}
 \hline Decay mode &$U$ & $\tilde{U}$ & $L$ & $\tilde{L}$ &$P$&$S$ & $\tilde{S}$&$\tilde{SL}$ & $\Gamma$ \\
 \hline${B^-_c}\to \eta_c e^-\nu_e$ & & & 4.844 & 4.432$10^{-7}$& & &15.397$10^{-7}$&4.712$10^{-7}$&4.844\\
 	  		
 \hline$B_c\to\eta_c \tau^-\nu_\tau$ & & & 0.756 & 0.172 && &1.194 &0.253&2.122\\
 	  		
 \hline${B^-_c}\to J/\psi e^-\nu_e$ &18.634&6.052$\times10^{-7}$ &16.283&27.813$\times 10^{-7}$&8.368 &1.188 &66.653$\times 10^{-7}$&22.856$\times 10^{-7}$&34.918\\
 	  		
 \hline${B^-_c}\to J/\psi \tau^-\nu_\tau$ & 3.823 &0.846 &1.922& 0.437&1.704 &0.614 &0.307&0.197&7.336\\
 	  		
 \hline${B^-_c}\to D e^-\nu_e$ &  &  &0.047&4.611$\times 10^{-10}$& & &1.072$\times 10^{-9}$&4.038$\times 10^{-10}$&0.047\\
 	  		
 \hline${B^-_c}\to D\tau^-\nu_\tau$ &  &&0.028&0.003& &&0.007&0.0027&0.038\\
 	  		
 \hline${B^-_c}\to D^* e^-\nu_e$ &0.2439&4$\times 10^{-9}$&0.078&7.760$\times 10^{-9}$&0.169&0.081&4.092$\times 10^{-8}$&3.648$\times 10^{-9}$&0.322\\
 	  		
 \hline${B^-_c}\to D^*\tau^-\nu_\tau$ &0.113&0.015&0.0156&0.0021&0.092&0.046&0.151&0.0094&0.297\\
 \hline
 \end{tabular}
 \end{table}    
In Table \ref{tab1}, we present the computed integrated partial helicity rates $\Gamma_i(i=U,L,P)$, $\tilde{\Gamma}_i(i=U,L,S,SL)$, and the total decay rates for the respective channels under study. We find the partial tilde rates for the $e^-$-mode to be insignificant and thus dismissable. However, the rates in the $\tau^-$-mode are comparable to the tilde rates and cannot be overlooked. Taking individual helicity rates into account, we provide decay rates predictions for each process in both $e^-$ and $\tau^-$ modes. Like all model predictions, our $\tau^-$ mode decay rates are typically smaller than those in the $e^-$ modes. For instance, our $B_c\to J/\psi$ transition decay rate prediction in $\tau^-$-mode is around 5 times smaller than its $e^-$ mode counterpart. For the $B_c\to \eta_c$ transition, the suppression factor is about 2. The $\tau^-$-modes for the CKM suppressed $B_c\to D(D^*)$ transitions are only slightly lower than their $e^-$-mode counterparts.
\subsection{Numerical analysis and results}
\label{subsec:Num}
Our aim is to examine LFU ratios, using the RIQM framework. We commence by assembling the input parameters used in this study. The relevant quark masses, the RIQ model parameters $(a, V_0)$, and quark binding energy $'E_q'$ as in \cite{Barik:1993aw,Barik:1997qq,Patnaik:2018sym,Patnaik:2017cbl} are determined from the model dynamics. The meson masses as, the $V_{\rm CKM}$, and the lifetime of the $B_c$ meson along with their uncertainities are taken from the PDG~\cite{ParticleDataGroup:2022pth}
\begin{eqnarray}
(a, V_0)&=&(0.017166\ {\rm GeV}^3,-0.1375\ {\rm GeV})\nonumber {\rm GeV}\\
m_b &\approx& 4.78 \,{\rm GeV}, \quad m_c \approx 1.5 \,{\rm GeV}, \quad m_u \approx 0.079 \,{\rm GeV}, \quad m_e \approx 0.51 \,{\rm MeV},\nonumber\\
m_{\tau} &\approx& 1.78 \,{\rm GeV}, \quad E_b \approx 4.77 \,{\rm GeV} \quad E_c \approx 1.58 \,{\rm GeV} \quad \ E_u \approx 0.47 \,{\rm GeV},\nonumber\\
m_{J/\psi} &\approx& 3096.9\pm 0.006\,{\rm MeV}, \quad \ m_{D} \approx 1864.84 \pm 0.05\,{\rm MeV}, \quad m_{B_c} \approx 6274.47 \pm 0.32\,{\rm MeV},\nonumber\\
m_{\eta_c} &\approx& 2983.9\pm 0.04\,{\rm MeV}, \quad m_{D^*} \approx 2006.85 \pm 0.05 \,{\rm MeV}, \nonumber\\
V_{cb} &\approx& 0.0408\pm0.0014, \quad \quad V_{bu} \approx 0.00382\pm0.00020, \quad \quad \tau_{B_c} \approx 0.51\pm 0.009\,{\rm ps}\,.
\end{eqnarray}
We compute the observable ${\cal R}$ using the model outlined in~\cite{Patnaik:2017cbl,Patnaik:2018sym}. The mechanisms behind the decay process are well comprehended within the context of a suitable phenomenological model (RIQ), achieved by comparing its predictions for observable phenomena with those from conventional theoretical methods. A quark potential model succeeds when it reasonably reproduces available observed data in different hadron sectors. Regardless of the Lorentz structure of the interacting potential used, a phenomenological model is considered reliable if it describes constituent-level dynamics within the hadron core and predicts various hadronic properties, including decays. However, the parameterization process at the potential level involves some arbitrariness. In this sense, the potential model approach is not unique, especially when limited to reproducing experimental data in a narrow range. Therefore, it is crucial to extend the applicability of a quark model to a broader range of observed data. In this scenarios, we have also adopted a potential model framework in the present analysis, which has been extensively used to describe a wide range of hadronic phenomena,

Our calculated ${\cal R}$ for $B_c\to X(nS) l \nu_l$, where $X = \eta_c , J/\psi$, $D,D^*$, in both ground and radially excited states, aligns well with other SM predictions as shown in Tables~\ref{table:2} and~\ref{table:3}. The mismatch between SM predictions of ${\mathcal R}$ and experimental data underscores anomalies in semileptonic decays and limitations of our RIQM in incorporating NP constraints. 

To quote here, in our current analysis, we have only focused on uncertainities stemming from meson masses, $V_{CKM}$ and lifetime of $B_c$. We have not incorporated uncertainities from the model framework in initial and final meson states, quark masses, corressponding binding energies  used in the calculation. It may be pointed out that, we do not have any adjustable free parameters in this model. The potential parameters ($a$, $V_0$), quark masses ($m_q$), corressponding binding energies ($E_q$) etc., have already been fixed from hadron spectroscopy in the basic level application of this model in reproducing the hyperfine splitting of baryons and mesons. We use the same set of fixed parameters to describe wide ranging hadronic phenomenas as pointed out earlier. As such we have no free parameters at our disposal to show any kind of uncertainties from our model framework. Therefore the dominant theoretical errors stem solely from the uncertainties of the input parameters $|V_{\rm cb}| = (42.2 \pm 0.8) \times 10^{-3}$, $|V_{\rm ub}| = (3.94 \pm 0.36) \times 10^{-3}$, $m_{\tau} = 1776.86 \pm 0.12$ MeV. The uncertainties in our predictions are concurrently within a range of $\pm 5\%$ of their central values. As a result, we have only reported the central values of our predictions in this work.

Given the lack of predictions from recognized models in existing literature, our predictions for LFUV observables for the higher charm and charmonium states, ${\cal R}_{D}(2S)$, ${\cal R}_{D}(3S)$, ${\cal R}_{D^*}(2S)$, and ${\cal R}_{D^*}(3S)$, may be instrumental in identifying the $B_c$ channels in the forthcoming Run 3 data at LHCb.
\begin{table*}[hbt!]
 \centering
 \setlength\tabcolsep{5.5pt}
 \renewcommand{\arraystretch}{2.0}
 \caption{Ratios of branching fractions for Semileptonic $B_c-$ decays in the ground state~\cite{Patnaik:2022moy}}
\resizebox{\textwidth}{!}{
\begin{tabular}{|c|c|c|c|c|c|}
 \hline Ratio of Branching fractions (${\cal R}$) ($l = e, \mu$) &RIQM&{(CQM)}\cite{Issadykov:2017wlb}&{RCQM}\cite{Ivanov:2006ni}&{PQCD}\cite{Wang:2012lrc} & LQCD \cite{Hu:2019qcn,Cooper:2021bkt}\\
 \hline ${\cal R}_{\eta_c}=\frac{{\cal B}(B_c\to \eta_c \tau \nu_{\tau})}{{\cal B}(B_c\to l \nu_l )}$&0.4325 &0.25&0.27&0.34 & - \\
 \hline ${\cal R}_{J/\psi}=\frac{{\cal B}(B_c\to J/\psi \tau \nu_{\tau})}{{\cal B}(B_c\to J/\psi l \nu_l )}$&0.21 & 0.23&0.236&0.29 & 0.27\\
 \hline${\cal R}_D=\frac{{\cal B}(B_c\to D \tau \nu_{\tau})}{{\cal B}(B_c\to D l \nu_l)}$&0.784&0.63&0.59&- &0.682\\
 \hline${\cal R}_{D^*}=\frac{{\cal B}(B_c\to D^* \tau \nu_{\tau})}{{\cal B}(B_c\to D^* l \nu_{l})}$&0.9165&0.56&0.58&-&-\\
 \hline
 \end{tabular}}
 \label{table:2}
 \end{table*}
 \begin{table*}[!hbt]
 \centering
 \setlength\tabcolsep{10pt}
 \renewcommand{\arraystretch}{2.0}
 \caption{LFU observables for $B_c$-decays to radially excited charmonium states~\cite{Patnaik:2022moy}}
 \resizebox{\textwidth}{!}{
 \begin{tabular}{|c|c|c|c|c|c|c|}
 \hline Ratio&RIQM&[BSA]\cite{Zhou:2020ijj}&[RQM]\cite{Ebert:2003cn,Ebert:2010zu}&[PQCD]\cite{rui2016semileptonic}&[LCQCD]\cite{Wang:2007fs}&[NRISGW2]\cite{Bediaga:2011cs}\\
 \hline ${\cal R}_{\eta_c}(2S)$&0.136&0.051&-&0.068&0.74&0.028\\
 \hline ${\cal R}_{\eta_c}(3S)$&0.021&0.0103&0.0009&0.0013&0.030&\\
 \hline ${\cal R}_{\psi}(2S)$&0.085&0.049&-&0.069&-&0.071\\
 \hline ${\cal R}_{\psi}(3S)$&0.077&0.009&0.0063&0.001&&\\
 \hline
 \end{tabular}}
 \label{table:3}
 \end{table*}
Our RIQM approach evaluates parametrization of relevant form factors for semileptonic decay amplitudes across the complete kinematic range ($0\le q^2\le q^2_{max}$), enhancing reliability and accuracy over other models. These models often calculate form factors with endpoint normalization at either $q^2 = 0$ or $q^2$ = $q^2_{max}$, then extrapolate them phenomenologically using monopoles, dipoles, and Gaussian ansatz to the entire physical region, potentially compromising reliability. To avoid such uncertainties, we refrain from using such phenomenological ansatz. Given the projected high statistics of $B_c$-events, expected to deliver up to $10^{10}$ events annually at colliders, semileptonic $B_c$ decays to charm and charmonium states offer an intriguing avenue to further explore emerging scales in physics.
\section{Tests of LFU violation in \texorpdfstring{$b \to s l^+ l^- $}{} transitions}
\label{sec:tests}
\begin{figure}[hbt!]
\centering
\includegraphics[width=\columnwidth]{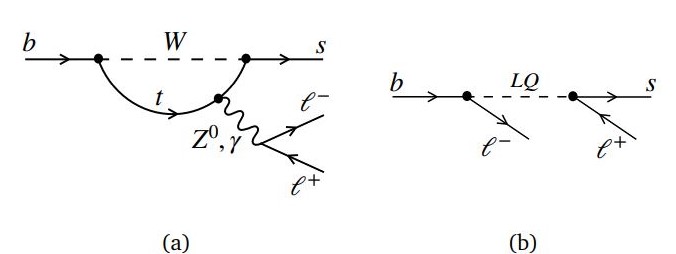}
\caption{(a) SM contribution for $b \to sl^+l^-$ transition including W, Z and $\gamma$ via loop level.
(b) NP contribution involving a leptoquark (LQ coupling directly to quarks and leptons)~\cite{Patnaik:2022moy}.}
\label{fig:1}       
\end{figure}
In the SM, quark flavor transitions are mediated solely by the charged weak bosons $W^{\pm}$. Hence, FCNC transitions involving same charge quarks aren't directly enabled by the neutral weak boson $Z^0$, but instead occur through rare loop processes involving virtual $W^{\pm}$ and extra virtual quarks. These transitions, depicted in penguin- and box-like Feynman diagrams, are accurately predicted by the SM. New particles can either partake in the loops or contribute additional tree-level diagrams, see Fig~\ref{fig:1}. Decays regulated by $b\to s l^+ l^-$ transitions, which are suppressed, are excellent environments to probe for new physics due to potentially sizeable effects compared to competing SM processes. 

Recently, significant deviations from SM predictions have been reported in LFU measurements related to these transitions. The authors of Ref.~\cite{LHCb:2021trn} demonstrated evidence of LFU violation with a 3.1$\sigma$ significance and around 5\% combined statistical and systematic uncertainty. Considerable interest has been drawn to decays at the quark level, like $b \to s l^+ l^- $, due to significant lepton universality violations. For the first time, these violations were reported by LHCb~\cite{LHCb:2014vgu, LHCb:2019hip, LHCb:2021trn}, with the result

\begin{equation}
	{\cal R}_K = 0.745({\rm stat}) \pm 0.036 ({\rm syst})\,,
\end{equation}
deviating from the SM prediction ${\cal R}_K = 1.0004$ \cite{Bordone:2016gaq} with a significance of 2.6\,$\sigma$.
Recent findings by LHCb have confirmed that ${\cal R}_K$ and ${\cal R}_{K^{*}}$ measurements align with SM predictions, showing no anomalies. The reported values are~\cite{LHCb:2022qnv}
\begin{align}
	{\cal R}_K &= 0.994({\rm stat}) \pm 0.029 ({\rm syst})\,, & {\cal R}_{K^{*}} &= 0.927({\rm stat}) \pm 0.036 ({\rm syst})\,, \\
	{\cal R}_K &= 0.949({\rm stat}) \pm 0.022 ({\rm syst})\,, & {\cal R}_{K^{*}} &= 1.027({\rm stat}) \pm 0.027 ({\rm syst})\,.
\end{align}
The measured ${\cal R}_K$ and ${\cal R}_{K^*}$ values for $q^2$ intervals: $0.1 < q^2 < 1.1~{\rm GeV}^2/c^4$ (low $q^2$ region) and $1.1 < q^2 < 6.0~{\rm GeV}^2/c^4$ (central $q^2$) supersede earlier measurements~\cite{LHCb:2021trn}, and are consistent with the SM. Data from all pp collisions recorded by the LHCb detector from 2011-2018 were used, equivalent to integrated luminosities of 1.0, 2.0, and 6.0 $fb^{-1}$ at 7, 8, and 13 TeV center-of-mass energies. Systematic uncertainties, significantly less than the statistical uncertainties, are projected to further reduce with additional data.

In the SM, decays with negligible electron-muon mass difference can be precisely predicted as they are exempted from hadronic uncertainties and only occur in loop order. Their suppression can be explained by SM's fundamental symmetries. 
Theoretical models~\cite{Isidori:2020acz} extensively examine universality effects, including QED-radiative corrections in the Monte Carlo framework, effectively negating non-perturbative QCD effects. Their work, using the meson effective theory's analytic results, identified ${\cal R}_K$ as a reliable SM prediction. Such corrections offer insight into the most sensitive moments and encourage experimental investigation. 

Since lepton universality violation observables can be influenced by the charm loop due to NP and SM contribution interference, an unbiased NP picture was investigated in~\cite{Ciuchini:2021smi} using charming penguins to solve charm loop amplitudes. Further details, including a detailed numerical comparison for neutral mode $\bar{B^0} \to \bar{K^0}l^+l^-$, which is relevant for LFU ratios, are in Ref.~\cite{Isidori:2022bzw}. However, Ref.~\cite{Altmannshofer:2021qrr} used an updated global fit of muon Wilson Coefficients to interpret anomalies. The significant 5.6\,$\sigma$ discrepancy motivates the proposition of various NP models.
The LHCb collaboration reported new measurements for ${\cal R}_{{K^0_s}}$ and ${\cal R}_{K^{*+}}$, and updated measurements for certain $B_s \to \phi \mu ^+ \mu ^-$ observables that deviate from the SM by a 3.6\,$\sigma$ level \cite{LHCb:2021lvy,LHCb:2021zwz,LHCb:2021xxq}. An effective field theory approach offered a solid data fit \cite{Alok:2022pjb}. 
Further analysis of $b \to s l^+ l^-$ decay transitions may provide insight into the observed matter-antimatter disparity in the universe. A study focusing on CP-violating angular observable with a complex phase offers a unique insight into potential NP structures in this transition~\cite{SinghChundawat:2022zdf}. 

Glashow, Guadagnoli, and Lane (GGL) proposed an explanation for the ${\cal R}_K$ anomaly, demonstrating that a NP model can explain both ${\cal R}_K$ and ${\cal R}_{D^{(*)}}$ anomalies using an effective field theory approach \cite{Bhattacharya:2014wla}. Under the assumption of NP predominantly coupling to the third generation and having a scale far beyond the weak scale, they identified two types of fully gauge-invariant NP operators, incorporating both neutral-current and charged-current interactions. Similar unified anomaly theories in the framework of effective theory are discussed in~\cite{Bhattacharya:2016mcc, Kumar:2018kmr, BhupalDev:2020zcy}.
\section{Conclusion \& future outlook}
\label{sec:conclusions}
Although the SM has been a potent mathematical framework, it's not without its limitations. This paper examines the recent landscape of LFUV anomalies in $B$ physics from both theoretical and experimental perspectives, which could unveil BSM physics. Notably, LFU ratios related to neutral-current (${\cal R}_K, {\cal R}_{K^*}$) align with SM predictions and are considered theoretically clean observables. 
Recently updated results for ${\cal R}_D$ and ${\cal R}_{D^*}$ show around 3\,$\sigma$ deviation, which has been explored in a model-dependent (RIQM) framework. NP models including leptoquark and others involving $Z^{\prime}$-boson~\cite{Mahata:2022cxf}, composite Higgs boson~\cite{Marzocca:2018wcf}, dark matter~\cite{Barman:2018jhz}, right-handed neutrinos~\cite{Mandal:2020htr} have been proposed to explain these discrepancies. If validated, these findings would provide clear evidence of NP.
The 2022-2025 LHC Run 3 is projected to triple data collections over three years, leading to improved event statistics, reduced uncertainties, and unprecedented precision for flavor measurements. Several upgrades to the LHCb detector will lower background from charged and neutral tracks and increase accessibility to electronic and tauonic modes. The Belle II analysis will also be instrumental in independently clarifying flavor anomalies. 

From a theoretical perspective, we should thoroughly report results from current models and encourage development of more NP models to explain these anomalies. A detailed examination of the upcoming Run 3 and its future implications has been reported in Ref.~\cite{Bernlochner:2021vlv}, which also discusses the Future Circular Hadron Collider (FCC-hh) at CERN's potential for direct observation of NP mediators up to the multi-TeV range.

Recent observations have indicated other anomalies suggestive of NP, including the anomalous magnetic moment of the muon and electron ($a_\mu$, $a_e$), and the mass of the W-boson. These could share an origin with anomalies in $B$ meson decays. The Fermilab (g-2) Collaboration's confirmation~\cite{Muong-2:2021ojo} of the old BNL result has increased the discrepancy from the data-driven SM prediction to about 4.2\,$\sigma$. Intriguingly, each of these flavor anomalies exceeds a 3\,$\sigma$ significance. The survival of at least one of these anomalies could herald a new era in our understanding of physics. If the LFUV anomalies persist, they may soon provide compelling evidence of NP within flavor physics, prompting intensive investigation from future experimentalists and theorists~\cite{ILCInternationalDevelopmentTeam:2022izu,Abir:2023fpo}.
\bmhead{Acknowledgments}
We thank Sanjay Swain and Sudhir Vempati for fruitful discussions.
R.S. acknowledges the support of Polish NAWA Bekker program no.: BPN/BEK/2021/1/00342 and Polish NCN Grant No. 2018/30/E/ST2/00432.\\

\noindent{\textbf{Data Availability Statement:} No Data associated in the manuscript.}
\bibliography{sn-bibliography}{}
\end{document}